\documentclass[english,aps,prl,superscriptaddress,floatfix,twocolumn,letterpaper]{revtex4}
\usepackage[T1]{fontenc}
\usepackage[latin9]{inputenc}
\usepackage{graphicx}
\usepackage{amssymb}
\usepackage{esint}

\makeatletter
\@ifundefined{textcolor}{}
{%
 \definecolor{BLACK}{gray}{0}
 \definecolor{WHITE}{gray}{1}
 \definecolor{RED}{rgb}{1,0,0}
 \definecolor{GREEN}{rgb}{0,1,0}
 \definecolor{BLUE}{rgb}{0,0,1}
 \definecolor{CYAN}{cmyk}{1,0,0,0}
 \definecolor{MAGENTA}{cmyk}{0,1,0,0}
 \definecolor{YELLOW}{cmyk}{0,0,1,0}
 }

\makeatother

\usepackage{babel}

\begin{document}

\title{Shards of Broken Symmetry: Topological Defects as Traces of the Phase
Transition Dynamics }

\thanks{This article is an updated version of the manuscript that has appeared
under the same title in the proceedings \cite{procLH} of the NATO
Advanced Study Institute held in Les Houches, France, in February
of 1999.}

\author{W. H. Zurek}

\address{Theoretical Division, Los Alamos National Laboratory, Los Alamos
NM 87545, USA }

\author{L. M. A. Bettencourt}

\address{Theoretical Division, Los Alamos National Laboratory, Los Alamos
NM 87545, USA }

\author{J. Dziarmaga}

\address{Theoretical Division, Los Alamos National Laboratory, Los Alamos
NM 87545, USA }

\author{N.D. Antunes}

\address{Dépt. de Physique Théorique, Université de Genéve, 24 quai E. Ansermet,
CH 1211, Genéve 4, Switzerland}
\begin{abstract}
We discuss the origin of topological defects in phase transitions
and analyze their role as a {}``diagnostic tool'' in the study of
the non-equilibrium dynamics of symmetry breaking. Homogeneous second
order phase transitions are the focus of our attention, but the same
paradigm is applied to the cross-over and inhomogeneous transitions.
The discrepancy between the experimental results in $^{3}He$ and
$^{4}He$ is discussed in the light of recent numerical studies. The
possible role of the Ginzburg regime in determining the vortex line
density for the case of a quench in $^{4}He$ is raised and tentatively
dismissed. The difference in the anticipated origin of the dominant
signal in the two ($^{3}He$ and $^{4}He$) cases is pointed out and
the resulting consequences for the subsequent decay of vorticity are
noted. The possibility of a significant discrepancy between the effective
field theory and (quantum) kinetic theory descriptions of the order
parameter is briefly touched upon, using atomic Bose-Einstein condensates
as an example. 
\end{abstract}
\maketitle

\section{Introduction}

The theory of the creation of topological defects appeals to models
of critical dynamics and to our understanding of the processes which
occur when phase transitions take place. Consequently, topological
defects can be used as {}``symptoms'', macroscopic manifestations
of underlying physical processes, which in turn can help diagnose
the nature of critical dynamics.

For first order phase transitions there is little doubt that nucleation
- a process understood for over half a century - is an essentially
accurate, universal yet simple model. A similarly simple model of
the dynamics of second order phase transitions was proposed much more
recently \cite{Z84,Z85,Z93}. One of its implications is the ability
to predict the size of the ordered patches of the new lower symmetry
phase, right from its inception. This allows one to calculate the
initial density of topological defects through the estimate put forward
in the seminal paper by Kibble \cite{K76,K80}. It may also lead to
a revision of the scenarios for baryogenesis and chiral symmetry restoration
\cite{Gill,DS}, as well as other related phenomena (such as the A-B
$^{3}He$ transition, see \cite{Cze,BT}). Some of the predictions
based on the new paradigm have been successfully tested and refined
in numerical experiments \cite{LZ97,LZ98,YZ98,ABZ,HS}. More importantly
the prediction of copious vortex production in superfluid phase transitions
has been experimentally verified in $^{3}He$ by two very different
strategies in two distinct parameter regimes \cite{He3G,He3H}. The
situation in $^{4}He$ \cite{He4new} and the initial indications
from high temperature superconductors \cite{supercond} are, however,
at best inconclusive. Indeed there are still differences concerning
analytic estimates of the initial density of defects in the underdamped
case \cite{Lythe,D98}, even in 1D. The aim of this paper is two-fold:
We shall start with a brief summary of the paradigm on which the emerging
understanding of second order phase transitions is based. We shall
then explore its extrapolations and investigate the experimental,
numerical and analytic evidence for and against this mechanism, in
various settings. This paper is not really an introductory survey
to the extent to which our lectures were. We have decided that the
existing literature (including the reviews of Zurek \cite{Z96} and
Eltsov, Krusius and Volovik \cite{EKV} as well as the other papers
mentioned above and below) already serves this purpose. Rather we
aim to perform a {}``reconnaissance by force'' of what is likely
to be the most interesting {}``proving grounds'' for the ideas summarized
briefly in the following section.

\section{Critical dynamics and defect formation}

Second order transitions fall into universality classes which are
characterized by the behavior of the healing length $\xi$ and the
relaxation time $\tau$ (among other quantities) as a function of
the relative temperature \begin{eqnarray}
\epsilon(T)=\frac{T-T_{c}}{T_{c}}.\end{eqnarray}
 Thus, $\tau\sim\vert\epsilon\vert^{-\nu z}$ and $\xi\sim\vert\epsilon\vert^{-\nu}$
diverge in the vicinity of $\epsilon=0$, where $\nu$ and $z$ are
universal critical exponents. A very specific model which represents
a large class of second order phase transitions is the so-called Landau-Ginzburg
theory. There, the dynamics of the order parameter is thought to effectively
obey a Langevin equation of the form: \begin{eqnarray}
\ddot{\varphi}+\eta\dot{\varphi}-c^{2}\nabla^{2}\varphi+\frac{1}{2}\left[\lambda\varphi^{3}+m^{2}\phi\right]={\cal O}(t,x).\label{Lang}\end{eqnarray}

Above, $\eta$ characterizes the viscosity in the system, while $c$,
$\lambda$ are constant coefficients. The mass term $m^{2}$ can depend
explicitly on time. Moreover, the correlation function for the noise
\begin{eqnarray}
\langle{\cal O}(t,x){\cal O}(t',x')\rangle=2\eta\Theta\delta(x-x')\delta(t-t'),\label{noise}\end{eqnarray}
 includes the temperature parameter $\Theta$, which can vary. The
change of $m^{2}$, eg. $m^{2}=m_{0}^{2}\epsilon(t)$ or of $\Theta$
or both, may precipitate the phase transition.

We shall assume that in the vicinity of the critical temperature $\epsilon(t)$
obeys a simple relation \begin{eqnarray}
\epsilon(t)=t/\tau_{Q}.\end{eqnarray}
 In that case the dynamics of the order parameter can be approximately
divided into the adiabatic and impulse regimes \cite{Z85,Z93}, with
the boundary which occurs at time $\hat{t}$ when the relaxation time
of the order parameter equals the characteristic time on which $\epsilon(t)$
changes: \begin{eqnarray}
\tau(\epsilon(\hat{t}))=\frac{\epsilon(\hat{t})}{\dot{\epsilon}(\hat{t})}={\hat{t}}.\end{eqnarray}

The timescale on which the order parameter will react to changes of
$\epsilon(t)$ depends on whether $\dot{\varphi}$ or $\ddot{\varphi}$
dominates. For the Landau-Ginzburg theory, in the two cases \begin{eqnarray}
\tau_{\dot{\varphi}}=\frac{\eta\tau_{0}^{2}}{\vert\epsilon\vert},\qquad\tau_{\ddot{\varphi}}=\frac{\eta\tau_{0}}{\vert\epsilon\vert^{2}},\end{eqnarray}
 respectively, where $\tau_{0}=1/m_{0}$, and $m_{0}$ is the mass
term in Eq.~(\ref{Lang}), evaluated for $T=0$ (i.e. for $\epsilon=-1$).

Using $\epsilon(t)=t/\tau_{Q}$, we can now solve for $\hat{t}$,
to obtain: \begin{eqnarray}
{\hat{t}}_{\dot{\varphi}}=\pm\tau_{0}\sqrt{\eta\tau_{Q}},\qquad{\hat{t}}_{\ddot{\varphi}}=\pm\tau_{0}^{2/3}\tau_{Q}^{1/3}.\end{eqnarray}
 To estimate the scale of the domains which could have become uniform
through dynamics in the adiabatic regime, we should need: \begin{eqnarray}
{\hat{\epsilon}}_{\dot{\varphi}}=\pm\sqrt{\frac{\eta\tau_{o}^{2}}{\tau_{Q}}};\qquad{\hat{\epsilon}}_{\ddot{\varphi}}=\pm\left(\frac{\eta\tau_{o}^{2}}{\tau_{Q}}\right)^{2/3}.\label{Z1}\end{eqnarray}
 The characteristic scale is then given by $\hat{\xi}=\xi_{0}/\sqrt{\vert{\hat{\epsilon}}\vert}$,
which yields: \begin{eqnarray}
{\hat{\xi}}_{\dot{\varphi}}=\xi_{0}\left(\frac{\tau_{Q}^{2}}{\eta\tau_{0}^{2}}\right)^{1/4};\qquad{\hat{\xi}}_{\ddot{\varphi}}=\xi_{0}\left(\frac{\tau_{Q}}{\tau_{0}}\right)^{1/3}.\label{Z2}\end{eqnarray}
 The initial density of defects can now be estimated using an argument
due to Kibble \cite{K76}, which will imply a $\hat{\xi}$-sized unit
of defect per $\hat{\xi}$ sized volume, i.e. $n\sim1/{\hat{\xi}}^{2}$
for vortex strings in two spatial dimensions.

The above calculation is based on the assumption that the order parameter
approximates its equilibrium configuration until $-\hat{t}$, at which
point it ceases to evolve dynamically (although noise and damping
continue unabated). The dynamical evolution restarts at $+\hat{t}$,
below the critical point, but by then it may be too late to undo non-trivial
topological arrangements of $\varphi$ inherited from above $T_{c}$.

This same paradigm decides when the overdamped or underdamped estimates
are relevant. For, in view of the above argument, it is essential
to decide whether the dynamics of the order parameter is overdamped
at $\hat{t}$, i.e. whether; \begin{eqnarray}
\eta{\dot{\varphi}}_{\vert{\hat{t}}}>{\ddot{\varphi}}_{\vert{\hat{t}}}.\end{eqnarray}
 This can be evaluated directly from Eq.~(\ref{Lang}) with the help
of the above estimates for $\hat{t}$ and $\hat{\epsilon}$, and leads
to the inequality: \begin{eqnarray}
\left(\eta\tau_{0}\right)^{3}>\tau_{0}/\tau_{Q}.\label{ineq}\end{eqnarray}

Numerical studies have by now confirmed this paradigm. The scalings
which we obtained follow theoretical predictions both when the quench
is induced by an explicit change of the mass term in Eq.~(\ref{Lang})
\cite{LZ97,LZ98}, \cite{YZ98} and when the temperature of the noise
$\Theta$ is adjusted, but $m$ set to a constant \cite{ABZ}. Moreover,
the switch from the overdamped to the underdamped behavior occurs
where expected, and with the consequences consistent with the scaling
implied by the paradigm Eqs.~(\ref{Z1})-(\ref{ineq}) \cite{LZ97,LZ98},
\cite{YZ98}. The same reasoning can be of course repeated using other
values of critical exponents relevant for other cases \cite{HH},
which has been already done in some cases \cite{Z96},\cite{ABZ}.

While the scalings accord well with the theoretical predictions, the
specific density of defects $n$ is lower than the appropriate inverse
power of $\hat{\xi}$; \begin{eqnarray}
n=\frac{1}{\left(f{\hat{\xi}}\right)^{2}},\end{eqnarray}
 where $f$ is always more than unity, and usually in the range $8-15$
\cite{LZ97,LZ98,YZ98,ABZ}. We shall return to its estimates later
in this paper.

\section{Crossover transitions}

An interesting case of transitions which does not conform to the second-order
universality class occurs when the critical scaling behavior of the
healing length and of the relaxation time {}``tapers off'' (i.e.
is fully analytic) very near to $\epsilon=0$. For instance a hypothetical
relaxation time and healing length dependences; \begin{eqnarray}
\tau=\frac{\tau_{0}}{\left(\vert\epsilon\vert+\Delta\right)};\qquad\xi=\frac{\xi_{0}}{\sqrt{\vert\epsilon\vert+\delta}},\end{eqnarray}
 illustrate such a crossover transition.

Examples of crossover phenomena are ubiquitous. One of the most interesting
cases comes from the study of the electroweak standard model, where,
for Higgs masses not yet excluded by experiment, the transition appears
to be a crossover \cite{SM}. A crossover transition may also occur
in the presence of impurities, anisotropies, weak external fields,
or finite size scaling, instead of the expected critical behavior
of the ideal model. It also substitutes critical behavior when non-perturbative
fluctuations exist in the spectrum of the theory which are favored
entropically and can destroy long range order. An example of the latter
is a $\lambda\phi^{4}$ theory or a (short-range) Ising model in one
spatial dimension.

Thus, $\tau$ and $\xi$ in the one-dimensional cases investigated
numerically \cite{LZ97,LZ98} as examples of the second order phase
transition are expected to taper off in the immediate vicinity of
the critical temperature. Presumably this occurs for very small values
of $\Delta$ and $\delta$, so that the scaling behavior encountered
in the vicinity of $\hat{\epsilon}$ is not affected. Nevertheless,
it is interesting to investigate what does the paradigm predict in
the case of such crossover transformations.

We follow the footsteps of the argument outlined in the preceding
section, and obtain $\hat{t}$ by solving : \begin{eqnarray}
\tau(\epsilon(\hat{t}))=\hat{t},\end{eqnarray}
 which now leads to the quadratic equation \begin{eqnarray}
\vert\hat{t}\vert^{2}+\vert\hat{t}\vert\Delta\tau_{Q}-\tau_{0}\tau_{Q}=0.\label{quadr}\end{eqnarray}
 Consequently \begin{eqnarray}
\vert\hat{t}\vert=\frac{-\Delta\tau_{Q}+\tau_{Q}\sqrt{\Delta^{2}+4\tau_{0}/\tau_{Q}}}{2},\end{eqnarray}
 and \begin{eqnarray}
\vert\hat{\epsilon}\vert=\frac{-\Delta}{2}+\frac{1}{2}\sqrt{\Delta^{2}+4\tau_{0}/\tau_{Q}},\end{eqnarray}
 where we have picked the physically relevant root of Eq.~(\ref{quadr}).

We note that in the limit of a `real' second order phase transition
($\Delta\rightarrow0$) we recover the old result, Eq.~(\ref{Z1}),
providing that the change of notation ($\tau_{0}$ now used to be
$\eta\tau_{0}^{2}$ in Eq.~(\ref{Z1})) is acknowledged. On the other
hand, when the quench is very slow and $\Delta^{2}>>4\tau_{0}/\tau_{Q}$,
$\epsilon\rightarrow\Delta\tau_{0}/\tau_{Q}$, which itself is small
compared with $\Delta$ (and, presumably also $\delta$ since $\Delta\sim\delta$
can be expected). Consequently for relatively rapid quenches \begin{eqnarray}
\hat{\xi}=\frac{\xi_{0}}{\sqrt{\left(\frac{\tau_{0}}{\tau_{Q}}\right)^{1/2}+\delta}},\end{eqnarray}
 which approaches Eq.~(\ref{Z2}) for sufficiently small $\delta$.
In the other limit; \begin{eqnarray}
\hat{\xi}=\xi_{0}/\sqrt{\delta},\end{eqnarray}
 and the size of the coherent domains of the order parameter saturates.

We note that the above discussion should be regarded more as an exercise
in extending the paradigm rather than as a generically valid theory,
applicable to all crossover phase transitions. In particular, in some
cases second order transitions may change into crossovers when an
external bias which influences the choices of the broken symmetry
vacuum is introduced. In such cases the externally imposed (rather
than spontaneous) symmetry breaking will favor a particular vacuum
and will lead to a suppression of topological defect production \cite{D98},
\cite{DS}).

Moreover, in case of the crossover transitions the influence of the
Ginzburg regime may need to be carefully examined as its role in the
generation and survival of topological defects is still a subject
of dispute.

\section{ Inhomogeneous Transitions }

Homogeneous quenches are a convenient idealization and may be a good
approximation in some cases. However, in reality, the change of thermodynamic
parameters is unlikely to be ideally uniform:\\
 1) Experiments carried out in $^{3}He$ \cite{He3G,He3H}, where
a small volume of superfluid is re-heated to normal state, and subsequently
rapidly cools to the temperature of the surrounding superfluid, are
a good example of an inhomogeneous quench: The normal region shrinks
from the outside. Yet, topological defects are created, thus suggesting
that the phases of distinct domains within the re-heated region are
selected independently. \\
 2) Another example are relativistic heavy ion collisions where,
according to Bjorken scenario \cite{bjorken}, a finite volume of
quark-gluon plasma can be created. The plasma expands in the direction
of collision and cools from the outside in the perpendicular direction.
The phase transition in this case can be first or second order (or
a smooth crossover) depending on the parameters of the collision.\\
 3) Any generic experiment based on pressure and/or temperature
quench is to some degree inhomogeneous because of finite velocity
of sound and finite heat conductance.

The mass parameter $\epsilon(t,\vec{r})$, varying in both time and
space, must be considered in defect formation. As a consequence, locations
entering the broken symmetry phase first could communicate their choice
of the new vacuum as the phase ordered region spreads in the wake
of the phase transition front. When this process dominates, symmetry
breaking in various, even distant, locations is no longer causally
independent. The domain where the phase transition occurred first
may impose its choice on the rest of the volume, thus suppressing
or even halting production of topological defects. This happens if
velocity of the critical front is less than certain characteristic
velocity.

\subsection{ Second Order Transition }

The characteristic velocity in an overdamped transition can be estimated
as follows: The freeze-out healing length is set at $\hat{t}$ as
$\hat{\xi}=\xi_{0}~(\tau_{Q}/\tau_{0})^{1/4}$. At the same instant
the relaxation time is $\hat{\tau}=(\tau_{Q}\tau_{0})^{1/2}$. These
two scales can be combined \cite{Z85} to give a velocity scale

\begin{equation}
\hat{v}=\hat{\xi}/\hat{\tau}=v_{0}~(\tau_{0}/\tau_{Q})^{1/4}\;\;,\end{equation}
 where $v_{0}=\xi_{0}/\tau_{0}$.

The density of defects $N$ as a function of critical front velocity
is expected to change qualitatively at $\hat{v}$. Above $\hat{v}$
the homogeneous estimates should hold. Below $\hat{v}$ the density
should be suppressed. Kibble and Volovik \cite{kibvol} suggested
that $N\sim v/\hat{v}$ for small $v<\hat{v}$. Dziarmaga, Laguna
and Zurek \cite{dlz} argued that $N$ is exponentially suppressed
below $\hat{v}$. There is qualitative difference between the two
proposals. The former option suggests that however one makes a quench
one will always get some defects, the latter implies that if one's
inhomogeneous quench is sufficiently slow one will get no defects
at all. In what follows we will quantify what {}``sufficiently slow''
means.

\paragraph{ Decay of the False Vacuum}

As a simple warm up exercise, let us consider decay of a false symmetric
vacuum to a true symmetry broken ground state in a one-dimensional
dissipative $\varphi^{4}$ model

\begin{equation}
\partial_{t}^{2}\varphi+\eta\,\partial_{t}\varphi\;-\;\partial_{x}^{2}\varphi\;+\frac{1}{2}\;(\varphi^{3}-\epsilon\varphi)=0\;,\label{varphi4}\end{equation}
 where $\varphi(t,x)$ is a real order parameter and $\epsilon$ measures
the degree of symmetry breaking i.e. $m^{2}=-\epsilon$. Without loosing
generality, we look for a solution $\varphi(t,x)$ which interpolates
between $\varphi(t,-\infty)=-\sqrt{\epsilon}$ and $\varphi(t,+\infty)=0$.
Such a solution can not be static. It is a stationary half-kink

\begin{equation}
\varphi(t,x)=-\sqrt{\epsilon}\left(1+\exp{\left[\frac{\sqrt{\epsilon}}{2}\frac{(x-v_{t}t)}{\sqrt{1-v_{t}^{2}}}\right]}\right)^{-1}\label{half-kink}\end{equation}
 moving with characteristic velocity \begin{equation}
v_{t}=\left[1+\left(\frac{2\eta}{3\sqrt{\epsilon}}\right)^{2}\right]^{-1/2}\stackrel{\eta\rightarrow\infty}{\approx}\frac{3\sqrt{\epsilon}}{2\eta}\;\;.\label{V}\end{equation}
 It is worth noting that the decay velocity $v_{t}$ increases with
$\epsilon$.

\paragraph{ Shock Wave }

Our shock wave inhomogeneous quench model consists of a sharp {}``pressure
front'' propagating with velocity $v$; that is,

\begin{equation}
\partial_{t}^{2}\varphi+\eta\,\partial_{t}\varphi\;-\;\partial_{x}^{2}\varphi\;+\frac{1}{2}\;(\varphi^{3}-\epsilon(t,x)\,\varphi)={\cal O}(t,x)\;\;,\label{model}\end{equation}
 where

\begin{equation}
\epsilon(t,x)\;=\mbox{Sign}(t-x/v)\label{a}\end{equation}
 is the relative temperature and ${\cal O}(t,x)$ is a Gaussian white
noise of temperature $\Theta$ with correlations given by Eq.~(\ref{noise}).

There are two qualitatively different regimes:

1) $v>v_{t}$, the phase front propagates faster than the false vacuum
can decay. The half-kink (\ref{half-kink}) lags behind the front
(\ref{a}); a supercooled symmetric phase grows with velocity $v-v_{t}$.
The supercooled phase cannot last for long; it is unstable, and the
noise makes it decay into the true vacuum.

2) $v<v_{t}$, the phase front is slow enough for a half-kink to move
in step with the front, $\varphi(t,x)=H_{v}(x-vt)$. The symmetric
vacuum decays into one definite non-symmetric vacua. The choice is
determined by the boundary condition at $x\rightarrow-\infty$. No
topological defects are produced in this regime. The stationary solution
$H_{v}(x-vt)$ is stable against small perturbations \cite{dlz}.

These expectations are borne out by the numerical study of kink formation
in ~\cite{dlz}. Numerical results are presented in Fig.~1.

\begin{figure}
\begin{centering}
\includegraphics[width=8cm]{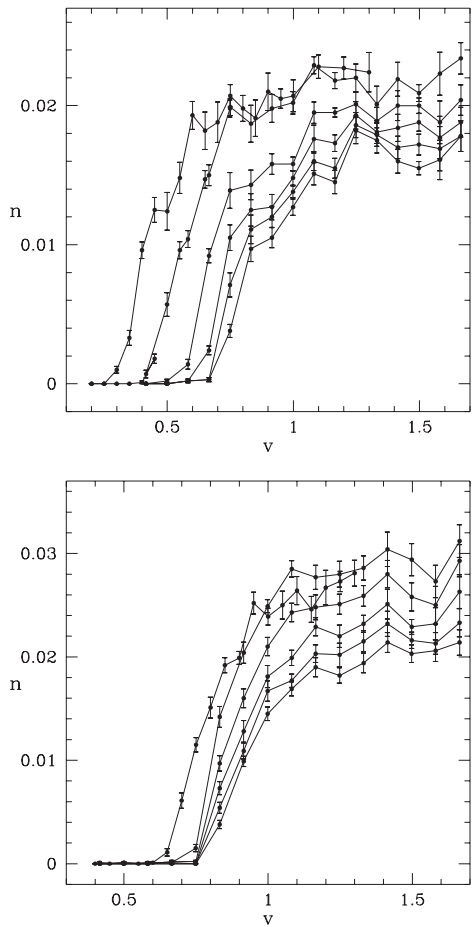}
\par\end{centering}

\caption{a. Density of kinks $n$ as a function of velocity $v$ for the shock
wave (\ref{a}) with $\eta=1$ (overdamped system). In this overdamped
regime, the predicted threshold velocity is $v_{t}=0.83$. The plots
from top to bottom correspond to $\Theta=10^{-1},\,10^{-2},\,10^{-4},\,10^{-6},\,10^{-8},\,10^{-10}$.
At low $\Theta$, we get a clear cut-off velocity at $v\approx0.8$,
which is consistent with the prediction. b. Density of kinks $n$
as a function of velocity $v$ for the linear inhomogeneous quench,
Eq.~(\ref{alinear}), with $\tau_{Q}=64$ and $\eta=1$. The predicted
threshold is $v_{t}=0.77$. This cut-off is achieved for low $\Theta$.
The plots from top to bottom correspond to $\Theta=10^{-1},\,10^{-2},\,10^{-4},\,10^{-6},\,10^{-8},\,10^{-10}$.
\label{shock}}

\end{figure}

\paragraph{ Linear Front}

Let us consider now a system in which the inhomogeneous quench takes
place via linear transition

\begin{equation}
\epsilon(t,x)=(t-x/v)/\tau_{Q}\label{alinear}\end{equation}

In the absence of noise, the propagating front is followed by a stationary
half-kink. This half-kink moves somewhat behind the front. Its location
is determined by the place where the threshold velocity (\ref{V})
is equal to the front velocity, $v_{t}[\epsilon(t,x)]=v$. The distance
between the front and the half-kink increases as $v^{3}$. This distance
gives the size of the supercooled region. When the supercooled region
is narrow then it is stable against small perturbations so that no
defects are produced. If

\begin{eqnarray}
v\;>\; v_{t} & \equiv & \left(1+\frac{\eta^{3/2}\tau_{Q}^{1/2}}{11.7}\right)^{-1/2} \nonumber \\
 & \stackrel{\eta\rightarrow\infty}{\approx} & \frac{3.42}{\eta}(\frac{\eta}{\tau_{Q}})^{1/4}\equiv\;4.07\;\hat{v}\;\;.\end{eqnarray}
 then the region is broad enough to be unstable \cite{dlz} and the
production of defects is no longer suppressed.

This prediction is confirmed by the numerical study of linear quenches
in Ref.~\cite{dlz}, compare Fig.~1. However, the threshold velocity
apparently gradually decreases with increasing noise temperature $\Theta$.
This decrease of the threshold for kink formation is due to the thermal
nucleation of kinks. Quantitative estimates for this effect are given
in \cite{kopnin}.

\subsection{ First Order Transition }

We assume the transition is strongly first order and that it goes
by bubble nucleation. To be more specific we consider a toy model
in 3 dimensions

\begin{equation}
\partial_{t}\varphi=\nabla^{2}\varphi-a\varphi+b\varphi^{3}-c\varphi^{5}+{\cal O}\;.\label{modelI}\end{equation}
 where $\varphi$ is real order parameter. The effective potential
is of the $\varphi^{6}$ type. Provided that $b^{2}>4ac$, it has
symmetric minimum at $\varphi=0$ and two symmetry broken minima at
$\varphi=\pm\varphi_{m}\equiv\pm\sqrt{(b+\sqrt{b^{2}-4ac})/2c}$.
We assume that $b,c$ are constant and that symmetry breaking transition
is driven by $a$ decreasing below its critical value $a_{c}=3b^{2}/16c$.
At $a=a_{c}$ all three minima are degenerate.

\paragraph{Decay of the False Vacuum }

Suppose that $a<a_{c}$. Let us consider decay of the false symmetric
vacuum to the true symmetry broken phase in a one dimensional version
of the model Eq.~(\ref{modelI}). We look for a solution which interpolates
between $\varphi=\varphi_{m}$ for $x\rightarrow-\infty$ and $\varphi=0$
for $x\rightarrow+\infty$. The solution is a stationary half-kink
$H(x-v_{t}t)$ moving with velocity

\begin{equation}
v_{t}=\frac{-b+2\sqrt{b^{2}-4ac}}{\sqrt{3c}}\end{equation}
 which has an envelope function

\begin{equation}
H(x)=\frac{\varphi_{m}}{\sqrt{1+\frac{\exp{\alpha x}}{2c}}}\;\;,\end{equation}
 where $\alpha=\sqrt{4c/3}\varphi_{m}^{2}$. This way the false $\varphi=0$
vacuum decays into the true $\varphi=\varphi_{m}$ vacuum in the absence
of noise. The decay velocity $v_{t}$ is zero for $a=a_{c}$, it increases
with increasing supercooling or with decreasing $a$.

\paragraph{ Shock Wave }

In the shock wave model a sharp front propagates with velocity $v$

\begin{equation}
a=a_{c}-\Delta a\;\mbox{Sign}(t-x/v)\;.\end{equation}
 Similarly as for second order transitions there are two regimes:

1) $v>v_{t}$, the pressure front propagates faster than the false
vacuum can decay. The half-kink lags behind the front. The supercooled
phase in between them grows linearly with time. The phase is unstable,
it decays by bubble nucleation just as for a homogeneous transition.
Homogeneous estimates of defect density apply in this case.

2) $v<v_{t}$, the half-kink is faster. It moves in step with the
front while its tail penetrating into the symmetric phase. There is
no supercooled phase where bubbles could be nucleated. The symmetric
phase goes smoothly into one of the symmetry broken phases.

\paragraph{ Linear Front }

Let the inhomogeneous quench proceed by a linear front moving with
velocity $v$

\begin{equation}
a=a_{c}-(t-x/v)/\tau_{Q}.\end{equation}
 The half-kink follows the critical front staying at a certain distance
behind it. The distance $D$ is such that the half-kink velocity $v_{t}$,
which depends on the local value of $a$, is equal to the front velocity
$v$, $v_{t}(a)=v$. With increasing $v$ the half-kink settles at
increasing values of local $a$. Close to the critical front the radius
of the critical bubble is infinite and at the same time the nucleation
rate is infinitely small. As we go away from the front in the direction
of the half-kink the critical radius shrinks. At a certain distance
$L$ from the front the energy of the critical bubble becomes comparable
to the temperature $\Theta$. At this point bubble nucleation becomes
possible. If $L<D$ bubbles can be nucleated in the supercooled region
between the front and the half-kink. If $L>D$ then there is no bubble
nucleation and no defects can be born in the supercooled area.

Now we estimate the critical velocity such that $L=D$. The half-kink
is located at such an $a$ that $v_{t}(a)=v$. $L=D$ providing that
for this $a$ the energy of the critical bubble $E(a)$ is equal to
temperature $\Theta$. The critical bubble is a metastable spherically
symmetric static solution of Eq.~(\ref{modelI}) with, say, $\varphi_{m}$
vacuum inside and $0$ vacuum outside its wall. Its energy can be
easily estimated when the width of its wall is negligible as compared
to its radius $R_{c}(a)$. An approximate solution is given by $H[r-R_{c}(a)]$,
where the critical radius is

\begin{equation}
R_{c}(a)=\frac{\sqrt{12c}}{-b+2\sqrt{b^{2}-4ac}}\;\;.\end{equation}

The energy of the critical bubble $E(a)$ has a negative volume contribution,
$(4\pi R_{c}^{3}/3)V(\varphi_{m})$, and a positive surface tension
term, $(4\pi R_{c}^{2})\int dx\;[H'(x)]^{2}$. When the solution of
$v_{t}(a)=v$ is put into $E(a)$ and then the equation $E(a)=\Theta$
is solved, one obtains a critical velocity

\begin{equation}
v_{cr}=\left(\frac{\pi b(3b^{2}-6bc+16c^{2})}{4c^{3}\Theta}\right)^{1/3}\label{vcr}\end{equation}
 for $L=D$. For $v>v_{cr}$ bubbles can nucleate in between the half-kink
and the front and thus the necessary condition for topological defects
production is satisfied.

The formula for $v_{cr}$, Eq.(\ref{vcr}), is still a crude lower
estimate for the critical velocity. In fact it is not sufficient to
nucleate some bubbles. Individual bubbles would coalesce with the
half-kink without any chance to trap any nontrivial winding number.
The bubbles should be nucleated in large numbers or have enough time
to grow so that they can mutually coalesce before merging with the
half-kink. Still, the argument which leads to $v_{cr}$ demonstrates
that there is a threshold velocity for defect formation.

\subsection{ Higher Dimensions }

The theory can be generalized to higher dimensions and to a complex
order parameter in a straightforward manner. Its major result is that
a subthreshold inhomogeneous quench does not produce any variation
of the order parameter in the direction normal to the front. This
excludes any possibility of production of vortex loops or closed membranes
entirely contained in the bulk, as well as of any pointlike defects.
Some extended defects can grow into the bulk provided their seeds
were created at this edge of the system where the symmetry was broken
first. In first approximation such, say, vortices grow into the bulk,
following the passing front, while keeping their direction normal
to the front. In the end we do not get any chaotic tangle of strings
and string loops but parallel {}``combed'' vortices. There are two
important perturbations to this {}``combed'' picture:

1) Thermal fluctuations make the strings look more random but without
backtracking and with string tension tending to smooth the small scale
fluctuations. The ends of the strings and antistrings at the critical
front are wandering around. Eventually an end of a string and of an
antistring may meet so that the strings join into a half-loop with
its both ends attached to the initial edge of the system. String tension
shrinks the half-loop to the edge where it unwinds.

2) A much more efficient factor to remove vortices from the bulk are
their mutual interactions. Global parallel string and antistring attract
one another so that their ends at the critical front do not seek each
other at random but tend to fuse in a deterministic way. This mechanism
makes the number of strings in the bulk decay with increasing distance
between the front and the initial edge.

The factors (1) and (2) lead to a picture in which the critical front
initially draws some parallel strings and antistrings from the edge,
then the strings recombine by joining ends and shrinking back to the
edge. In the end only the net surplus of strings (or antistrings)
is left in the bulk.

These ideas are supported by experiments:

1) Disclinations produced during a quench from disordered to nematic
phase in liquid crystals \cite{LiqCr}. This is a weakly first order
transition. In early attempts to make cosmological experiments in
liquid crystals the disclinations were observed to grow approximately
combed, join ends and shrink to the initial edge. Later on it was
realized that these quenches were not homogeneous enough \cite{yurke}.

2) Czochralski method of growing monocrystals, which is widely used
to grow silicon monocrystals necessary for microchips. In this method,
discovered in the thirties, a surface of liquid material is touched
with a monocrystal template. As the template is slowly lifted up it
drags a column of crystal out of the container. The top part of the
column is cold while its bottom part is at the melting temperature
- the transition is inhomogeneous. If the template is lifted slowly
enough, then no defects of the crystal lattice are produced which
might spoil the monocrystal.

To conclude this section: in an inhomogeneous quench there is a threshold
velocity $v_{t}$ of the critical front. Above the threshold defects
are produced like in a homogeneous quench. Below the threshold one
gets no defects; instead a clean monocrystal or a {}``disoriented
chiral condensate'' is grown with a vacuum which may be uniform over
significant distances, but which differs from the true vacuum.

\section{Defect formation and the Ginzburg Regime}

Recently a new $^{4}He$ experiment \cite{He4new} was devised, improving
on the apparatus used earlier by McClintock \textit{et al}. \cite{He4Old}
to implement a superfluid transition in $^{4}He$ through a sudden
pressure quench. The corresponding results are rather surprising.
They show no evidence for the formation of topological defects at
the anticipated levels, contrary to expectations based both on the
old experiment \cite{He4Old}, the theory%
\footnote{Although a factor $f\stackrel{>}{\sim}10$ in the formula for the
string density $n\sim1/(f\hat{\xi})^{2}$ could explain the new results
and seems consistent with recent numerical studies \cite{ABZ}.%
} and the $^{3}He$ data \cite{He3G,He3H}. The discrepancy with the
earlier $^{4}He$ quench data is now seen as the evidence of mechanical
stirring in the first version of the experiment. Nevertheless to address
the discrepancy with $^{3}He$ it was suggested \cite{KR} that because
the Ginzburg regime in $^{4}He$ extends over a broad range of temperatures
around the $\lambda$-line, large scale fluctuations may be able to
unwind and alter the configuration of the order parameter (in contrast
to $^{3}He$) while the quench proceeds.

The Ginzburg temperature is defined through the loss of ability of
the order parameter to hop, through thermal activation, over the potential
barrier between broken symmetry vacua. Thus one might worry with Karra
and Rivers \cite{KR} that when the defect densities are eventually
measured, at a much later time, little or no string would have survived
unwinding through thermal activation. In this section we investigate
this possibility and more generally report a numerical study of the
effect of thermal fluctuations on topological defect formation and
evolution.

Originally the Ginzburg temperature $T_{G}$ was suggested to be the
time of formation of topological defects \cite{K76}, since, at lower
temperatures, thermal fluctuations would be unable to overcome the
potential energy barrier associated with the defect's topological
stability.

In reality the situation is more complex. In equilibrium at any given
temperature $T$ (including of course temperatures in the Ginzburg
regime) a range of string configurations will exist. However, long
strings can only exist in equilibrium strictly above $T_{c}$ \cite{AB}.

To freeze them out, i.e., to \textit{form} them, energy (associated
with the string tension) must be extracted from the system. This necessarily
breaks time invariance and will lead to locally preferred nonequilibrium
field configurations. Subsequently the system will order over larger
and larger spatial scales, leading to mutual string annihilation.

The initial density of defects entering this stage of evolution is
computed by the theory of section II. This density is set at an effective
temperature $-\hat{\epsilon}$, which in $^{4}He$ is well within
the estimates for the width of the Ginzburg regime. By contrast, in
$^{3}He$ the Ginzburg temperature is small compared to the typical
$\hat{\epsilon}$. What happens to the initial densities of string
when the system is exposed to temperatures in the Ginzburg regime
for an extended amount of time?

In order to investigate this issue we need a quantitative definition
of $T_{G}$. In tune with the arguments given above consider a volume
of characteristic size $\xi(T)$, the correlation length, and a theory
with two energetically degenerate minima of an effective potential
$V(\phi)$, separated by a potential barrier $\Delta V$. The rate
for the field to change coherently from one minimum to the other per
unit volume due to thermal activation is $\exp\left[-\Delta V/k_{B}T\right]$.
For an effective potential of the form (obtained, eg. perturbatively
at 1-loop) \begin{eqnarray}
V(\phi)=-\frac{1}{2}m^{2}(T)\varphi^{2}+\lambda\varphi^{4}\label{Veff}\end{eqnarray}
 $\Delta V=\frac{m(T)^{4}}{4\lambda}$. For a volume $\xi^{3}$, we
define $T_{G}$ such that the probability of overcoming the potential
barrier is of order unity: \begin{eqnarray}
T_{G}\ :\ \frac{\Delta V(T_{G})}{T_{G}}.\xi^{3}(T_{G})=1\ \ \Leftrightarrow\ \ \lambda T_{G}/m(T_{G})=\frac{1}{4} \ .\ \ \ \label{Gzg}\end{eqnarray}
 This definition however has some caveats, for instance, an effective
potential of the form Eq.~(\ref{Veff}) is only valid for the mean
field and not on smaller scales. A more careful accounting of scales
leads to different results \cite{Bettencourt}, which show an enhancement
of the hoping probability. Thus, the factor of $1/4$ in Eq.~(\ref{Gzg})
should not be taken at face value.

A more rigorous definition arises from the range of temperatures below
$T_{c}$ for which fluctuations are large and consequently where perturbative
finite temperature field theory fails to be useful. In order to set
up a perturbative scheme at finite temperature from an initial 3+1
dimensional quantum field theory one implements dimensional reduction
which is valid provided the temperature is high compared to all mass
scales. As a consequence the coupling of the dimensionally reduced
3D field theory becomes dimensionful, i.e. $\lambda\rightarrow\lambda T=\lambda_{3}$.
In order to proceed one has to identify an appropriate dimensionless
coupling. This is done by taking $\lambda T/m(T)$. The Ginzburg regime
is entered when this 3D effective coupling becomes strong, in the
vicinity of the critical point, namely \begin{eqnarray}
T_{G}\ :\ \lambda T_{G}/m(T_{G})=1.\label{Gz}\end{eqnarray}
 To compute $T_{G}$ one needs the scaling of $m(T)$ in the critical
domain. We write $m^{2}(T)=m_{0}^{2}\epsilon^{\nu}$, with $\epsilon$
being the reduced temperature $\epsilon=\vert\frac{T-T_{c}}{Tc}\vert$.

Thus $\epsilon_{G}=-0.18$ for $\nu=0.5$. This mean-field estimate
produces an upper bound in $T$ for $T_{G}$ (and lower bound for
$\beta=1/T$). For realistic 3D exponents, $\nu=0.67$, we obtain
$\epsilon_{G}=-0.25$. The first criterion, based on the hopping of
a correlation sized volume, results in higher values of $T_{G}$.
This brings about a relatively large uncertainty in the value of $T_{G}$,
which is $18-25\%$ below $T_{c}$.

\subsection{Strings Survive the Ginzburg Regime}

In order to investigate the role of the Ginzburg temperature in the
\textit{dynamics} of defect formation we deliberately expose the system
to a heat bath at temperature $\epsilon_{i}$, within the Ginzburg
regime and below. We repeat this procedure for a range of time intervals
$\Delta t$, after which the bath temperature is taken to zero. This
set of temperature trajectories is shown in Fig.~2. We are attempting
to emulate the worst case scenario of an experimental quench where
the temperature or pressure are dropped monotonically but where the
system makes a long stopover within the Ginzburg regime.%
\begin{figure}
\begin{centering}
\includegraphics[width=8cm]{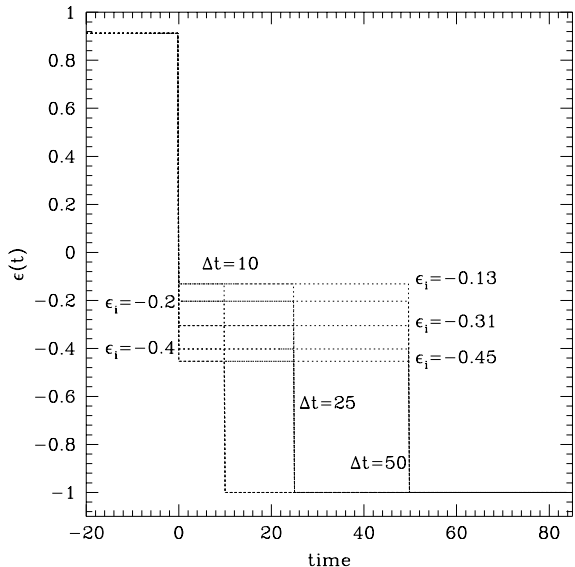}
\par\end{centering}

\caption{Temperature trajectories for testing the effect of exposure to the
Ginzburg regime on string densities. The system is first thermalized
at a high temperature and then placed in contact with a heat bath
at an intermediate temperature $\epsilon_{i}$ below $T_{c}$, for
a time interval $\Delta t$. \label{figtraj}}

\end{figure}

We would expect that, if the Ginzburg regime indeed produced enhanced
decay of strings, then the string densities measured at later times
should be smaller the longer the time the system spent within the
range $T_{c}\geq T\geq T_{G}$.

We have measured the final string densities at a time $t>>\Delta t$.%
\begin{figure}
\begin{centering}
\includegraphics[width=8cm]{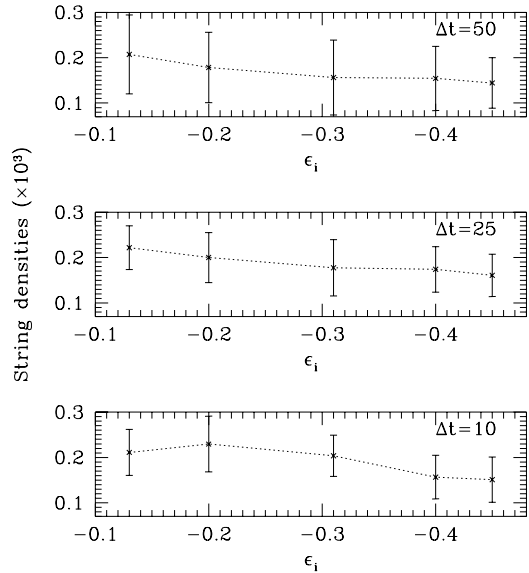}
\par\end{centering}

\caption{The string density measured at a later time $t>>\Delta t$ vs. intermediate
temperature $\epsilon_{i}$. From top to bottom the three plots correspond
to $\Delta t=10,20,50$, during which the system remained in contact
with a heat bath at $T_{i}$. There is no visible role played by intermediate
temperatures within the Ginzburg regime. \label{figGinz}}

\end{figure}

Our results for the final string densities as a function of intermediate
temperature $\epsilon_{i}$ and $\Delta t$ are summarized in Figure~3.
There is no apparent effect of the Ginzburg regime in reducing string
densities at formation.

If any trend is visible from Figure~3 it is the opposite, namely
that the lower $\epsilon_{i}$, the less string is measured at later
times. This is consistent with the relaxation of the string network,
resulting in vortex annihilation controlled by the string tension
(which is smaller near $T_{c}$) and with results the thermodynamics
of vortex strings \cite{AB}.

\subsection{Memory of the Order Parameter Configuration near $T_{c}$}

An independent test on the possible role of thermal fluctuations in
affecting string densities consists in reheating a quenched system
to a temperature around $T_{c}$ (both below and above it) and cooling
it again. This process tests the memory of the order parameter as
well as that of other related quantities {[}see also \cite{YZ98}{]},
such as defects. These temperature ($\epsilon(t)$) trajectories are
illustrated in Fig.~4a.

\begin{figure}
\begin{centering}
\includegraphics[width=8cm]{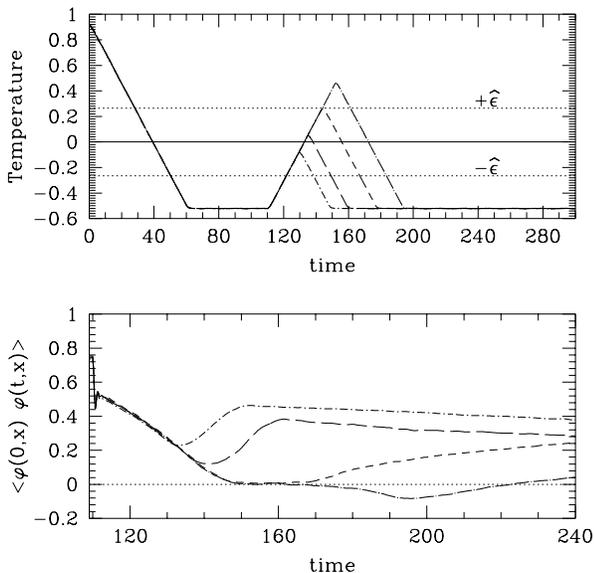}
\par\end{centering}

\caption{a. Dependence of the bath temperature $\epsilon$ in time. After being
quenched in temperature ($\tau_{Q}=80$) the system is reheated at
the same rate to a temperature $\epsilon_{f}=0.469,0.256,0.061,-0.068$
(top to bottom) `and cooled again. b. The correlation function between
the field at the time just before reheating and at later times, $\langle\varphi_{i}(t_{{\rm rh}},x)\varphi_{i}(t+t_{{\rm rh}},x)\rangle$
is plotted. There is a universal short time transient for the decorrelation
of the field over small scales while the long time tails of the correlation
function describe change over the mean fields. All four trajectories
cross the Ginzburg regime, but only those reaching or crossing $+{\hat{\epsilon}}$
display a significant memory loss. \label{fig4}}

\end{figure}

We are particularly interested in investigating under what circumstances
thermal fluctuations can affect the large scale configuration of the
order parameter.

We define the unequal-time correlation function \begin{eqnarray}
 \langle\varphi(x,t_{{\rm rh}})\varphi(x,t+t_{{\rm rh}})\rangle = \nonumber \\
 N\sum_{j=1}^{2}\sum_{i}\varphi_{j}(x_{i},t_{{\rm rh}})\varphi_{j}(x_{i},t+t_{{\rm rh}}),  \label{cor}\end{eqnarray}
 where $N$ is an irrelevant normalization factor. Note that $\varphi(x,t_{{\rm rh}})\varphi(x,t+t_{{\rm rh}})\rangle$
cannot be complex as we have summed over the field's components. This
correlator has several interesting properties. For short times it
displays a characteristic time, which describes the decay of correlations
over very small spatial scales. This is the initial transient in Fig.~4b.
For later times the residual correlation comes from the motion of
the order parameter (the field volume average). This average can be
either positive or negative but, if thermal, will converge to zero
at and above $T_{c}$.

Now, we are interested in determining whether the final field configuration
over large spatial scales is correlated to the configuration prior
to reheating. Fig.~4 shows that only if one crosses $T_{c}$, by
more than $+\hat{\epsilon}$, is the memory of the initial quenched
configuration erased (see in particular the two trajectories reaching
higher temperatures in comparison to the others). For these trajectories
the field correlations reach zero and after reheating evolve to a
value manifestly different from that prior to reheating.

For trajectories within the Ginzburg regime, that do not cross $T_{c}$,
the change in the configuration of the order parameter as measured
by Eq.~(\ref{cor}) is small. In particular the field configuration
existing before reheating is approximately recovered as the fields
are cooled. The same is true for the string densities, including those
of long strings.

Thus we are led to conclude (see also \cite{AB}) that even prolonged
exposure of a quenched field configuration to the Ginzburg regime
has little consequences in changing the order parameter configurations
emerging at $-\hat{\epsilon}$, and associated string densities. In
addition we have shown that to truly destroy a quenched field configuration
existing below $-\hat{\epsilon}$, one has to expose the system to
temperatures well above $T_{c}$. In particular for any particular
quench trajectory, a temperature of $T\sim T_{c}+\hat{\epsilon}$,
must be reached and maintained for a time $\sim\hat{t}$ in order
to erase memory of the initial configuration.

These results fully support the theory of section II for the critical
dynamics of second order transitions and all known thermodynamic results
for vortex strings in $O(N)$ theories. Thus we expect the results
of this section to carry over from our models to the Lancaster $^{4}He$
experiments. The results of Ref. \cite{He4new} in these experiments
cannot therefore be attributed to the effects of Ginzburg regime in
$^{4}He$.

In the next section we offer an alternative explanation.

\section{What is being observed in the $^{3}$He and $^{4}$He experiments.}

The several experiments in Helium, and more recently in superconductors,
testing the theory of defect formation rely on substantially different
processes to induce the phase transition and measure defects.

In this section we analyze, in the light of our own theoretical results,
how experimental procedures can lead to the detection of substantially
different defect densities.

Two particular factors play a decisive role in the value of the topological
defect density measured - the time and procedure of measurement after
the quench and the initial/final state of the system.

\subsection{The Lancaster experiments in $^{4}He$}

In the Lancaster experiments in $^{4}He$ the defect density is measured
through the attenuation of a second sound signal (a heat pulse). This
probe can only detect densities above a certain threshold (if the
theory of section II is used $f\leq10$ would be required, which is
at odds with the results of the numerical studies \cite{LZ97,YZ98}
and especially \cite{ABZ}). Moreover, the density at formation has
to be extrapolated from the data obtained at relatively late times
- the signal is noisy shortly after the quench \cite{He4new}.

After being formed by the critical dynamics of the phase transition
vortex strings decay away, as the system orders and cools. This decay
has been modeled by Vinen's Equation \begin{eqnarray}
\dot{n}=-\gamma n^{2},\end{eqnarray}
 where $n$ is the length of string per unit volume, i.e. the string
length density, $\gamma>0$. This model has been observed in the same
experiment to describe very well the decay of vorticity induced initially
through a fluid flow.

Vorticity created thermally is potentially different from that formed
under an external flow. We know from several theoretical and numerical
indications that a thermal distribution of vortices close to the transition
is comprised of both long strings and small loops, see Fig.~5.%
\begin{figure}
\begin{centering}
\includegraphics[width=4cm]{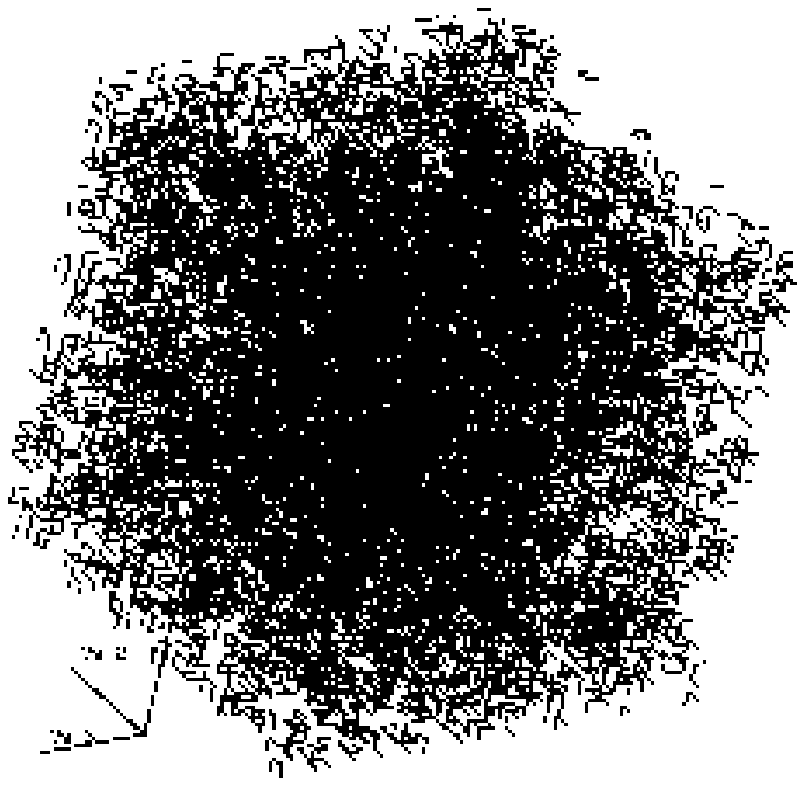}\includegraphics[width=4cm]{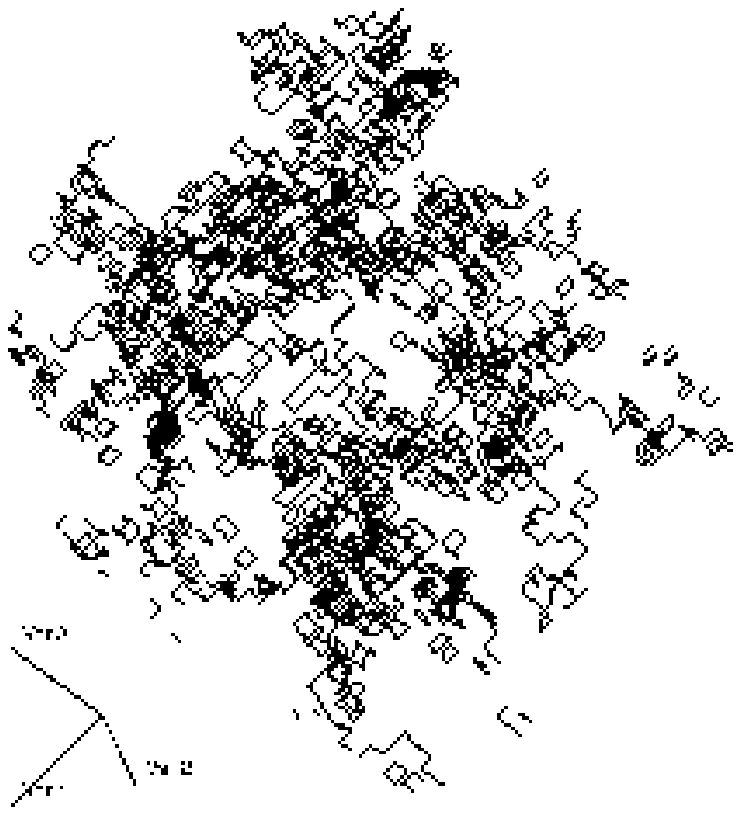}
\par\end{centering}

\includegraphics[width=4cm]{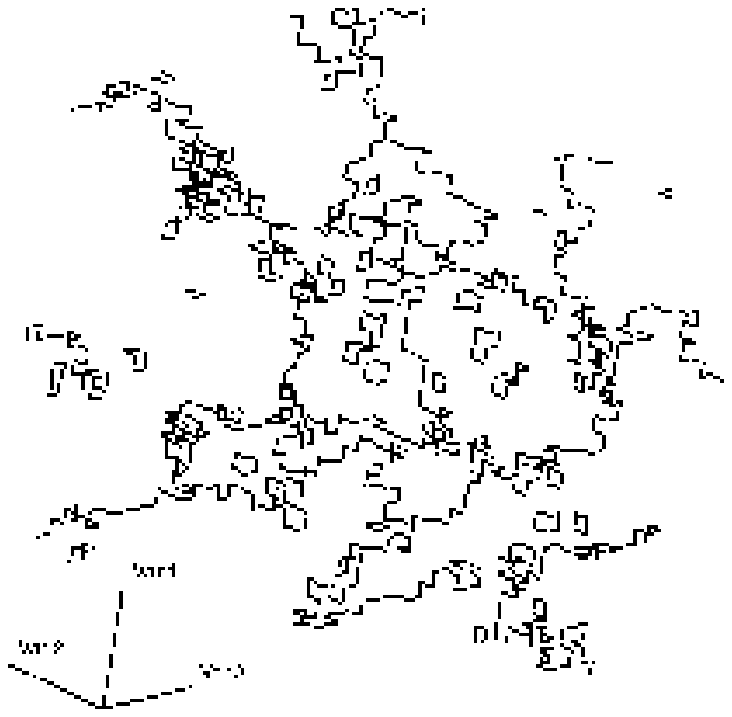}

\caption{The decay of vorticity under a quench. Initially the defect network
includes both long strings and small loops. At late times the network
coarsens and only long strings remain, see also Fig.~6. \label{fig6}}

\end{figure}

These two populations decay very differently in the wake of the quench.
Without any mechanism for stabilization the loops tend to disappear
in a fast transient. In contrast the long strings loose some of their
small scale structure but survive, and will ultimately set the decay
pattern described by the Vinen equation.%
\begin{figure}
\begin{centering}
\includegraphics[width=8cm]{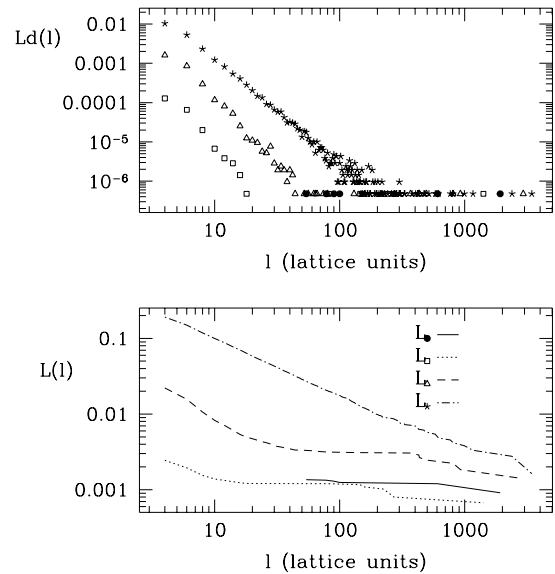}
\par\end{centering}

\caption{String length $l$ distributions Ld(l), taken between $+\hat{t}$
and the {}``time of formation'' ($\langle\vert\phi\vert\rangle=0.95$),
for $\tau_{Q}=64$. Data sets denoted by ($\star$,$\triangle$,$\Box$,
$\bullet$) correspond to increasingly later times. Lines show the
integral distributions, eg. $L_{\bullet}(l)=\sum_{l}^{\infty}l'Ld_{\bullet}(l')dl'$.
It is clear that at late times only long strings survive.\label{fig7}}

\end{figure}

It is the surviving long strings - eventually measured - that will
provide the experimental signal in the $^{4}He$ experiments. This
is shown in Figs.~5 and 6. As the system is quenched from higher
temperatures or pressures, an initial string network comprised of
strings of all lengths looses its loops and settles to a much slower
decay trend dominated by long strings.

The crucial question then is whether enough long string would persist
at the time of measurement to yield a positive signal.

We have performed a very similar procedure in a numerical {}``experiment''
\cite{ABZ}. We observed that at long times string densities could
be measured that agreed extremely well with all features of the theory.
Our definition of \textit{long} times was intimately connected to
the completion of the phase transition expressed in the expectation
values of the order parameter $\langle\vert\phi\vert\rangle\simeq0.9-0.975$.
The effective $f$ measured then was in the range $f=11-16$. All
our indications are that the Lancaster $^{4}He$ experiment performs
its measurements much later (up to 2 orders of magnitude) than we
do, thus leading to even smaller string densities. Such string densities
could evade detection under the second sound experimental probe, which
may lead to the negative result \cite{He4new}.

The other possibility is that the annihilation of a network of vortices
arising from a rapid quench may proceed at a rate different than for
the network produced by turbulent flow. Clearly, in case of turbulent
flow there may be correlation between the orientation of nearby vortices.
By contrast, vortices created by quench are anticorrelated (see, e.g.,
discussion in section 4 of Ref.\cite{Z96}). It seems plausible that
annihilation of vortices would be faster in that case. In order to
measure a positive signal in these circumstances the measurements
would have to be made sooner, after a much faster quench, or with
a higher sensitivity.

\subsection{Experiments in $^{3}He$}

In contrast to the experiments in $^{4}He$ described above which
appear sensitive to the {}``infinite'' string, experiments in $^{3}He$
have pursued two independent strategies both of which allow one to
stabilize and detect defect loops of various sizes: either vorticity
is stabilized and amplified by the flow and then measured directly
using nuclear magnetic resonance (the Helsinki experiment) or it is
inferred from a certain amount of missing energy (the Grenoble and
Lancaster $^{3}He$ experiments). Of these two procedures the first
is more direct - vorticity formed during the quench is forced to migrate
to the center of the container, through the existence of a subcritical
rotation velocity, where it is detected. This permits loops of string
of length larger than a known threshold to survive decay and results
naturally in a higher density, i.e. the defect density is measured
effectively very shortly after the transition takes place and need
not be limited the the {}``infinitely long string''. As a consequence,
much smaller values of $f$ and larger string densities are measured
than in the $^{4}He$ experimental setting.

Both remaining $^{3}He$ experiments end at a region of the phase
diagram far from the transition - $^{3}He$ being very much colder
than in the Helsinki experiment. A lower effective temperature results
in the effective absence of damping mechanisms which in turn leads
to the preservation of even small vortex loops. Dissipation mechanisms
rely on the presence of quasiparticles. Thus, when the medium is very
cold, energy dissipation will slow exponentially and vortices can
be stabilized by a coherent flow resulting from their motion through
the superfluid. We expect therefore the string population in all $^{3}He$
experiments to be mostly in the form of relatively small loops. In
Helsinki, the largest loops are stabilized by the slow rotation of
the whole system, and their density can be extrapolated to the smaller
loops, leading to the total consistent with the Grenoble and Lancaster
experiments, where - one may guess - all of the loops survive for
a long time in the absence of dissipation.

The lifetime of these loops is thus expected to be much longer than
that of thermal loops formed at a quench through the $\lambda$-line
in $^{4}He$. As a result the long time decay of vorticity may also
be very different in these two cases as the former corresponds to
an ensemble of moving loops, at relative distances much larger than
their typical radius, but the latter contains strings of all sizes,
where the mean distance between strings is comparable to their length.

This conjectured picture, supported in part by numerical studies,
leads to the conclusion that both experimental settings in $^{3}He$
should lead to a positive result, compatible with a relatively small
value of $f$ relevant for all loops (we get $f\simeq4$, when $n\sim1/({\hat{\xi}}f)^{2}$
is used to fit early data in Fig.~6), whereas in $^{4}He$ the smallness
of the signal at the time of first measurement makes the detection
more difficult and at present below the sensitivity threshold.

\section{Discussion}

The mechanism we described early on in this paper is based on the
analysis of the behavior of the order parameter $\varphi$. The order
parameter is clearly a phenomenological entity and the equations that
govern its evolution are approximate and in many cases postulated
rather than derived. On the other hand the underlying physics is usually
very specific. It may, for instance, involve atoms of some particular
isotope such as $^{4}He$. Thus, in principle, one could formulate
an exact microscopic theory of particular second order phase transformations.
However, in all of the experimentally accessible cases discussed above
such a fundamental theory is simply too complicated to lead to useful
conclusions. The superfluid transition in $^{4}He$ is a good example:
Strong interactions in $^{4}He$ make it impossible to proceed rigorously
all the way starting at the microscopic level. Analysis of related
issues in the field theoretic context is also difficult \cite{BVH}.
Recently however a new system has become experimentally accessible:
Atomic Bose-Einstein condensates (BEC's) undergo the second-order
phase transition at much lower densities. Natural approximation schemes
can be therefore suggested, and the exact microscopic theory can be
studied in greater detail than for the {}``old'' superfluids.

We shall not attempt to review the theoretical or experimental situation
in BEC's. Good reviews already exist (see eg. \cite{BECs}). Our aim
is simply to point out that questions concerning the formation of
topological defects can be posed and analyzed within a much more fundamental
formalism, which is explicitly quantum. The approximations start from
the Schrödinger, equation and lead in a controlled manner to master
equations for the density operator of the condensing system. Further
approximations result in a quantum kinetic theory. Preliminary analysis
of these issues \cite{AZ} allows one to recover the key scaling relations
and the key predictions we have described in section II. Indeed, time
dependent Landau-Ginzburg theory follows as an approximation to some
of the terms which one obtains from the microscopic treatments. On
the other hand, the microscopic theory contains additional terms,
which alter predictions concerning the formation of topological defects.
Limited studies \cite{AZ} indicate that the predicted densities of
the vortex lines or of the winding numbers would be smaller than those
based on the scalings of the order parameter (see section II of this
paper). Moreover, corrections seem to be more significant as the ratio
$\tau_{0}/\tau_{Q}$ decreases.

The possibility of experimental studies of defect formation in BEC
quenches nevertheless exists and may lead to exciting insights into
the problem.

Superconductors may be the other useful testing ground. Indeed, two
experiments have been already reported \cite{supercond}, with the
claim of conflicting results, which seemed to depend on the geometry.
Rapid cooling produced no detectable signal in a high-temperature
film, although it is far from clear whether any was expected. The
original claim that the effect was ruled out at the {}``$\sim10^{3}$
level'' was based on an overly optimistic prediction, which did not
recognize that the total flux expected to arise in the experimentally
studied geometry scales as $n^{1/4}$, i.e. only with the \textit{fourth
root} of the total number of defects (rather than with the square
root; see section 4 of Ref.\cite{Z96} for discussion).

The revised prediction is close to the claimed sensitivity of the
experiment, and given the uncertainties in the critical exponents
of the high-temperature superconductor, as well as the possibility
of imperfect trapping of the defects, etc. it is unfortunately impossible
to extract useful constraints from the existing negative experiment.

The experiment carried out by the same group, in the loop geometry
has, on the other hand, yielded positive results. This experiment
also operates near the edge of detectability. It detects the flux
induced by a loop which is artificially broken into a large number
N of superconducting sections, which are then rapidly reconnected.
The predicted flux should have a Gaussian distribution with a random
direction and intensity corresponding to $\sigma\sim\sqrt{N}$ flux
quanta \cite{Z96}, such signals appear to have been indeed found
\cite{supercond}.

The available experimental results can be therefore described as confusing.
In liquid crystals the results seem to be the perhaps least ambiguous,
but they concern a (weakly) first order transition. In superfluids,
$^{3}He$ is still the strongest case for the mechanism, especially
since the results between all the experiments (carried out in quite
different parameter regimes, and using very different techniques)
are consistent. On the other hand the relevance of the Helsinki experiment
for the cosmological scenario has been recently questioned by numerical
experiments \cite{aravin} which indicate that the vorticity generated
in such settings may be induced by the flow imposed in the Helsinki
experiment to facilitate the process of their detection. These simulations
were carried out under a very idealized set of assumptions (which
included very small fluctuations and axial symmetry) and conclusions
of Ref.\cite{aravin} appear to be inconsistent with the experiment
\cite{volaravin}, but much more remains to be done to clarify this
issue. Indeed, a fully 3-D study with large fluctuations is under
way \cite{3D}. Further experimental and numerical studies to investigate
the role of rotation in stabilizing vortex loops and to explore the
implications for the A-B phase transition, etc., are nevertheless
essential.

The existing $^{4}He$ data are clearly disappointing, but not at
odds with a more conservative theoretical estimate. Moreover as we
have argued above there may be a way to reconcile estimates of vortex
line density obtained from $^{3}He$ and $^{4}He$ experiments, even
without any special appeal to the Ginzburg regime \cite{KR} or to
quantum kinetic theory.

Finally the first experimental reconnaissance into quenches in superconductors
is preliminary in its nature and ambiguous in its results.

In the meantime, numerical studies have confirmed and refined the
basic indications of the theory of order parameter dynamics.

One may be therefore justified in the expectation of an exciting but
uncertain future. A lot is at stake, including the understanding of
the phase transition dynamics, nature of the order parameter and other
collective observables of quantum many body systems and perhaps even
the relation between the quantum and the classical.

\end{document}